\documentclass{aa}

\usepackage{txfonts}
\usepackage{graphicx}
\usepackage{natbib}
\usepackage[colorlinks=true,citecolor=blue]{hyperref}
\newcommand*{\mk}{ }

\newcommand{\orcid}[1]{\href{https://orcid.org/#1}{\includegraphics[width=10pt]{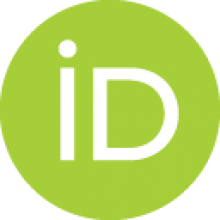}}}

\begin{document}

\title{Curvature Mapping Method: Mapping Lorentz Force in Orion A}

\author{Mengke Zhao\orcid{[0000-0003-0596-6608]}\inst{1,2,3}
\and Guang-Xing Li\orcid{[0000-0003-3144-1952]}\inst{2}
\and Keping Qiu\orcid{[0000-0002-5093-5088]}\inst{1,3}
}

\authorrunning{Zhao, Li, $\And$ Qiu}

\institute{
School of Astronomy and Space Science, Nanjing University, 163 Xianlin Avenue, Nanjing 210023, PR China \label{inst1} \\
\and South-Western Institute for Astronomy Research, Yunnan University, Kunming 650091, People’s Republic of China \label{inst2}\email{gxli@ynu.edu.cn} \\ 
\and Key Laboratory of Modern Astronomy and Astrophysics (Nanjing University), Ministry of Education, Nanjing 210023, Jiangsu, PR China \label{inst3}\email{kpqiu@nju.edu.cn}
} 

\abstract{
Magnetic forces are critical in regulating the star formation process, yet spatially resolving their dynamical role remains challenging. 
Traditional diagnostics like the mass-to-flux ratio measure field strength but lack directional information on how the force functions. 
We propose the Curvature Mapping Method (CMM) to reconstruct the 2D projected Lorentz force ($\boldsymbol{f}_{\rm L}$) vector field from polarization observations.
The method assumes that the magnetic field is deformed by external forces such as gravity, marking a transition from a force-free to a forced regime.
We validate the method using MHD simulations, confirming its reliability specifically in dense molecular cloud cores where the magnetic field structure is well-resolved and magnetic forces are dynamically significant.
We apply CMM to the OMC-1 region and compare the derived $\boldsymbol{f}_{\rm L}$ map with a spatially resolved gravitational force map ($\boldsymbol{f}_{\rm G}$) calculated via the Poisson equation. 
We identify a clear bimodal behavior: in the diffuse envelope ($\log N_{\rm H_2} \lesssim 23$ cm$^{-2}$), the magnetic force is statistically uncorrelated with gravity; however, along the dense filamentary ridge, $\boldsymbol{f}_{\rm L}$ vectors become systematically anti-parallel to gravity. 
The magnetic support ratio, {${-\boldsymbol{f}_{\rm L} \cdot \boldsymbol{f}_{\rm G}}/{|\boldsymbol{f}_{\rm G}|^2}$}, approaches unity in the filament backbone, providing direct evidence that magnetic fields provide substantial support against gravitational collapse, thereby regulating the fragmentation and star formation rate. 
The method effectively uses information contained in polarization maps and can be applied to data from surveys to understand the role of the B-field.
}

\maketitle

\section{Introduction}
The magnetic field plays an important role in {the} interstellar medium (ISM) and molecular clouds \citep{1981MNRAS.194..809L,2007ARA&A..45..565M,2021Galax...9...41L},
whose role in the star formation process is not entirely understood \citep{2012ARA&A..50...29C,2021Galax...9...41L,2023ASPC..534..193P}. 
Thermal dust polarization, arising from aspherical dust grains aligned with magnetic field lines \citep{2016A&A...586A.138P,2017ApJ...842...66W,2018JAI.....740008H}, is the main magnetic field tracer, which can constrain the magnetic field morphology in the plane of the sky (POS).
Large radio arrays, such as SMA and ALMA, provide high-resolution dust polarization data to study magnetic fields of the interstellar medium (ISM) in star formation \citep{2020ApJ...895..142L,2024arXiv240303437L}. 
The next stage is to analyze the information in these data sets to quantify the dynamical properties of magnetic fields, such as their field strength and their role in counteracting self-gravity. 
Various techniques have been proposed to measure magnetic fields in the ISM, such as the Zeeman effect and Faraday rotation \citep{2012ARA&A..50...29C,2025ApJ...978L..31R}.
However, for dust polarization observations in molecular clouds, the Davis-Chandrasekhar–Fermi (DCF; \citealt{1951PhRv...81..890D,1953ApJ...118..113C}) method is the most commonly used method, where the magnetic field strength is estimated through the dispersion of the polarization angle. 
The method is based on the assumption that the field lines are straight where the angle dispersion of the field lines is caused by turbulence, and this angle dispersion can be estimated using the dispersion of the polarization angle. 
However, in many cases, the magnetic field exhibits large-scale coherent structures \citep{2006Sci...313..812G,2014ApJ...794L..18Q,2015Natur.520..518L} where the assumption of almost-straight fields as required in the DCF breaks down. 
The question of how to analyze these data sets becomes an urgent one.

In contrast to the DCF method, which attributes magnetic field distortions to turbulent motions, the curvature of ordered field lines serves a different dynamical role: generating magnetic tension to counterbalance gravity. 
In fact, those pinched magnetic field structures are usually interpreted as a balance between magnetic force and gravity in observations \citep{2015Natur.520..518L,2017ApJ...846..122P,2021A&A...647A..78A}. 
Following this reasoning, \cite{2015Natur.520..518L} used this force balance to estimate the magnetic field strength by the balance between gravity and the Lorentz force. 
However, previous studies \citep{1998ApJ...493..811S,2012ApJ...747...79K,2014ApJ...794L..18Q,2015Natur.520..518L} only applied this method to some particular cases.
We argue that this argument is general and can estimate the Lorentz force in a much larger variety of situations compared to \citealt{2015Natur.520..518L}.  
In this paper, we make the working assumption that  the observed curvature arises from the resistance of the magnetic field to external compression (e.g., self-gravity), moving the system away from a force-free state ($\mathbf{J} \times \mathbf{B} = 0$) toward a dynamic equilibrium.
The 2D magnetic field, as traced by e.g. the dust polarization observations, reflects the projection of the 3D magnetic field, through which the projected component of the magnetic force on the plane of sky can be estimated through the equation:
\begin{equation}
    f_{\rm L, POS} \approx \Omega f_{\rm t, 2D} = \frac{\Omega}{\mu_0} {\rm B}^2\cdot \boldsymbol{\kappa} \,,
\end{equation}
where {$B$ and $\boldsymbol{\kappa}$} are the strength scalar and curvature vector of magnetic field, {and $\Omega$ is a numerical factor.}
This approach involves two underlying assumptions: First, the Lorentz force is comparable to the magnetic tension ($f_{\rm L} \approx f_{\rm t}$). Second, the 2D Lorentz force measured in our method reflects the 2D projected component of the 3D Lorentz force. 
As the discussion of the paper unfolds, one should see that both assumptions hold to some reasonable accuracy. Thus, one can estimate the behavior of the magnetic force using the dust polarization observations, which opens up new approach to the direct analysis of the effect of the magnetic force in the interstellar gas. 

\begin{figure*}
    \centering
    \includegraphics[width = 14cm]{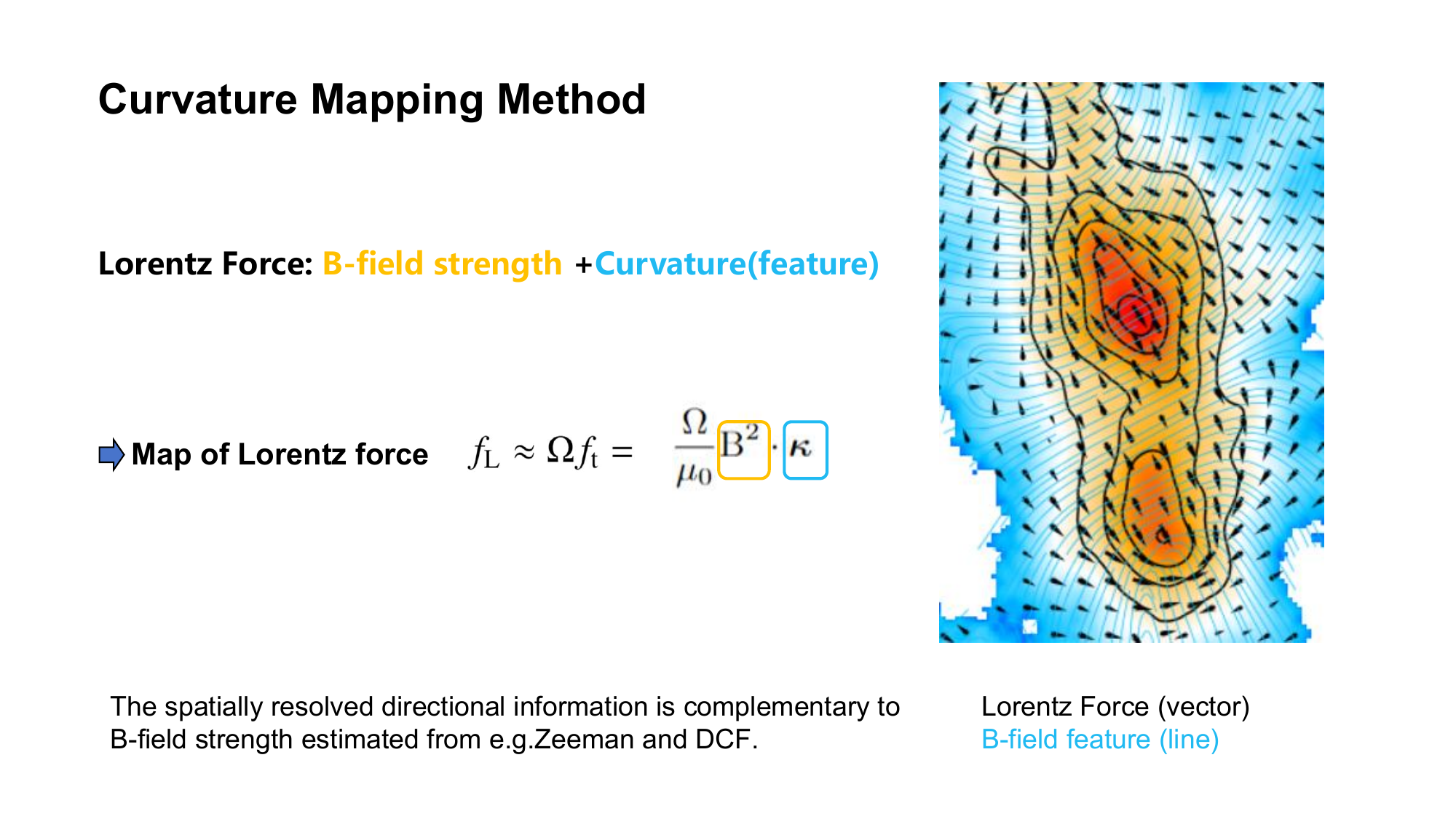}
    \caption{{\bf Curvature Mapping method: mapping the Lorentz force in molecular clouds.}
    Magnetic field strength and the curvature of the field lines combined to derive the Lorentz force
    Previous studies estimate magnetic field strength using Zeeman splitting or the DCF method.
    In comparison, the Curvature Mapping Method enables the inference of spatially resolved maps of the magnetic force from polarization observations.
    The right panel displays a zoomed-in region of OMC-1 (see Fig.\ref{figcurvature}) to illustrate the vector map construction.
    }
    \label{figcarton}
\end{figure*}

\section{Curvature Mapping Method}\label{method}

\subsection{{Magnetic Tension as a proxy for the Lorentz Force}}\label{2.1}

The Lorentz force is defined as the force acting on moving charged particles in an electromagnetic field and is related to the magnetic field $\boldsymbol{B}$ and current density $\boldsymbol{J}$ in plasma:
\begin{equation}
    f_L = \boldsymbol{J} \times \boldsymbol{B} \,,
\end{equation}
Using Ampère's law ($\boldsymbol{J} = (\nabla \times \boldsymbol{B})/\mu_0$), the Lorentz force can be expressed in terms of the magnetic field as:
\begin{equation}\label{eqf_L}
    f_L = \frac{1}{\mu_0} (\nabla \times \boldsymbol{B}) \times \boldsymbol{B} = \frac{(\boldsymbol{B}\cdot \nabla)\boldsymbol{B}}{\mu_0} - \nabla\left(\frac{\boldsymbol{B}^2}{2\mu_0}\right) \,,
\end{equation}
where the first term represents magnetic tension and the second term represents magnetic pressure force.


Magnetic tension, acting as the restoring force against field line bending, is a fundamental component of the Lorentz force.
Direct comparisons between magnetic tension and the Lorentz force in MHD numerical simulations using the Enzo code \citep{2012ApJ...750...13C,2015ApJ...808...48B} reveal that the magnetic tension closely resembles the Lorentz force in both magnitude and orientation (details provided in Section~\ref{Ap.B}). 
While magnetic pressure force contributes significantly to the force magnitude (comparable to tension), MHD simulations (see Sect. \,\ref{Ap.B}) reveal that the magnetic tension vector aligns much more closely with the total Lorentz force orientation in dense gas than the pressure gradient does. 
This factor accounts for the combined effects of the magnetic pressure force and the 3D-to-2D projection geometry, bridging the gap between the projected tension and the total 3D Lorentz force.
The Lorentz force is estimated as:
\begin{equation}\label{eqf_t}
    f_L \approx \Omega_1 f_t = \frac{\Omega_1}{\mu_0}(\boldsymbol{B}\cdot \nabla)\boldsymbol{B} =\Omega_1 \frac{{\rm B}^2}{\mu_0}(\Vec{b}\cdot \nabla)\Vec{b} = \Omega_1\frac{{\rm B}^2}{\mu_0} \cdot \boldsymbol{\kappa} \,,
\end{equation}
where ${\rm B}$ represents the magnetic field strength, $\Vec{b}$ is the unit vector of the magnetic field, and $\boldsymbol{\kappa}$ is the magnetic field curvature vector derived from the magnetic field morphology.
Since the magnetic tension, pressure force and Lorentz force are similar in the simulation (see Sect.\,\ref{Ap.B}), the weighting factor $\Omega$ is introduced to account for the contribution of the magnetic pressure force (which is not directly traced by curvature) to the total Lorentz force magnitude.
{The Tension Ansatz:} using tension to represent the {Lorentz} force works
in the kinematic regime, but not the force-free region, and this corresponds to
the high-density part of the ISM \citep{2025MNRAS.542.3246L} where the magnetic 
force is forced rather than force-free.
This assumption allows the Lorentz force to be decomposed into two components: a scalar parameter representing magnetic field strength and a vector parameter derived from the magnetic field morphology. 

\subsection{{Observational Adaptation: From 3D to 2D}}

The observation of the magnetic field, different from data of the MHD simulation, only has the 2D magnetic field orientation \citep[e.g., polarization]{2017ApJ...846..122P,2019ApJ...872..187C,2020A&A...641A..12P} in the plane of sky (POS) and the estimation of total magnetic strength \citep[e.g., DCF method, Zeeman splitting]{2023ASPC..534..193P,2010ApJ...725..466C}.
Thus, we assume the 2D magnetic tension is close to the Projected component of Lorentz force to measure the projected Lorentz force on POS.
The robustness of this assumption is quantitatively verified in Section~\ref{sect3.3}, where we demonstrate that the derived 2D force orientation closely tracks the true projected Lorentz force in dense environments. 
According to Eq.\,\ref{eqf_t}, this projected Lorentz force depends on the magnetic curvature on POS.

{The 2D curvature vector} can be credibly calculated by POS magnetic field orientation in most cases.
Curvature is the key parameter of our method, which is calculated by the unit vector of the magnetic field.
The POS unit vector can be obtained from magnetic field orientation $\phi$, the x, y components as the cos$\phi$ and sin$\phi$. 
To trace the projected component of the Lorentz force, the magnetic field orientation observed in the POS via dust polarization is assumed to represent the main structure of the 3D magnetic field. 
Assuming a random distribution of magnetic field orientations \citep{2004ApJ...600..279C}, the probability of the offset angle ($\theta$) between the 3D field vector and the POS being less than 45$^\circ$ is approximately 71$\%$ (based on the probability density $p(\theta) \propto \sin\theta$). 
This implies that in most cases, the POS structure captures the dominant magnetic field morphology.
However, projection effects inherently introduce uncertainty into individual magnitude estimates, which scale with $1/\cos\theta$. 
To address this, we quantitatively evaluated the impact of projection in Section~\ref{sect3.3}. 
As shown in Fig.~\ref{fig3}, we analyzed the ratio of the derived force to the true 3D Lorentz force. 
The results demonstrate that in dense gas, this ratio distribution is well-constrained. 
This confirms that the statistical correction factor, $\Omega$, effectively accounts for the projection geometry.
The POS magnetic field structure could present the main magnetic field morphology to calculate the projected curvature and Lorentz force in most cases.
 The components of the magnetic curvature vector $\boldsymbol{\kappa}$ can then be calculated as:
\begin{equation}\label{eqkx}
    \kappa_x = {\rm cos}\phi_B \frac{\partial }{\partial{x}}{\rm cos}\phi_B + {\rm sin}\phi_B\frac{\partial }{\partial{y}}{\rm cos}\phi_B \,,
\end{equation}
\begin{equation}\label{eqky}
    \kappa_y  = {\rm cos}\phi_B \frac{\partial }{\partial{x}}{\rm sin}\phi_B + {\rm sin}\phi_B\frac{\partial }{\partial{y}}{\rm sin}\phi_B \,,
\end{equation}
\begin{equation}\label{eqkxy}
    \boldsymbol{\kappa} = \kappa_x \boldsymbol{i} + \kappa_y \boldsymbol{j} \,,
\end{equation}
where $\kappa_x$ and $\kappa_y$ are the curvature components along the x- and y-axes.

\subsection{{Tracing the Projected Lorentz Force}}

The projected component of Lorentz force in the plane of the sky (POS) can be measured using the POS magnetic field morphology and magnetic field strength (see Fig.\,\ref{figcarton}). 
For easy application in observation, the Projected Lorentz force is calculated by the magnetic curvature {and total magnetic field strength in the 2D plane} 
\begin{equation}\label{eqcurvforce}
    f_{\rm L, projected} \approx {\rm C} f_{\rm t, 2D} = {\rm C}\frac{\Omega_1}{\mu_0}B_{\rm POS}^2 \boldsymbol{\kappa} = \frac{\Omega}{\mu_0}B_{\rm tot}^2 \boldsymbol{\kappa} \,,
\end{equation}
where $\Omega$ is a modified factor including projection effects of the Lorentz force, the effect of magnetic pressure force, typically ranging between 1 and 1.5 (detailed in Sect.\,\ref{sec3.3.2}).

The POS-projected Lorentz force thus depends on two primary physical parameters: the scalar term $B_{\rm tot}$ (magnetic field strength) and the vector term $\boldsymbol{\kappa}$ (magnetic field curvature). 
The curvature vector $\boldsymbol{\kappa}$ can be derived from magnetic field orientations, generally observable through dust polarization \citep{2017ApJ...842...66W,2020A&A...641A...1P}. 
The total Magnetic field strength, $B_{\rm tot}$, can be estimated using many methods such as Zeeman splitting \citep{2010ApJ...725..466C}, the DCF method \citep{1951PhRv...81..890D,1953ApJ...118..113C,2004ApJ...600..279C}, the ST method \citep{2021A&A...647A.186S,2021A&A...656A.118S}, and so on.

As advances in observations, the projected Lorentz force can be mapped as vectors in the POS, including both its strength and orientation. 
This technique, the *Curvature Mapping Method*, has the added advantage of resolving the Lorentz force orientation over the full range of $0$ to $2\pi$.
Importantly, curvature is invariant under magnetic field vector direction, ensuring:
\begin{equation}
    (\Vec{b}\cdot \nabla)\Vec{b} = (\Vec{-b}\cdot \nabla)\Vec{-b} \,.
\end{equation}
Thus, the curvature orientation spans $0$ to $2\pi$, unaffected by the polarization-based magnetic field orientation ($0$ to $\pi$) observed in the POS.
Caution is required in its application under certain conditions, such as when the magnetic field structure is perfectly straight without curved or when magnetic pressure force effects dominate significantly, since the assumptions of this method may be less valid in these cases, although these scenarios could be infrequent \citep{2016A&A...586A.138P,2021ApJ...909..148G}.
In most cases, the Curvature Mapping Method effectively measures the projected Lorentz force using the POS magnetic field morphology.
The verification of this method is presented in the next section.

\section{{Verification via MHD Simulations}}

\subsection{Simulations}

The numerical simulation of molecular cloud selected in this work is identified within the three $\beta_0$ simulations applied by the constrained transport MHD option in Enzo (MHDCT) code \citep{2010ApJS..186..308C,2015ApJ...808...48B}. 
The simulation conducted in this study analyzed the impact of self-gravity and magnetic fields on supersonic turbulence in isothermal molecular clouds, using high-resolution simulations and adaptive mesh refinement techniques. which detail in shown in \citep{2012ApJ...750...13C,2015ApJ...808...48B,2020ApJ...905...14B}. 
The Enzo simulation with three dimensionless parameters provides us with massive information on physical processes:
\begin{equation}
    {\cal M}_s = \frac{v_{\rm rms}}{c_s} = 9
\end{equation}
\begin{equation}
    \alpha_{vir} = \frac{5 v_{\rm rms}^2 }{ 3 G \rho_0 L_0^2} = 1
\end{equation}
\begin{equation}\label{eq3}
    \beta_0 = \frac{8\pi c_s^2 \rho_0}{B_0^2} = 0.2, 2, 20
\end{equation}
where $v_{\rm rms}$ is the rms velocity fluctuation, c$_s$ is sonic speed, $\rho_0$ is mean density, and $B_0$ is the mean magnetic field strength.
With the same sonic Mach number ${\cal M}_{\rm s}$ and viral parameter $\alpha_{vir}$ in initial conditions, the difference only exists in initial magnetic pressure force $\beta_0$ with various B-field strength,
The $\beta_0$ as 0.2, 2, and 20 present the strong B-field state, medium B-field, and weak B-field state.
The clouds are in three different evolutionary stages started with self-gravity
existing at 0.6 $t_{\rm ff}$ {($\sim 0.76 $ Myr)}, where $t_{\rm ff}$
represents 1.1$((n_{H} / 10^3)^{-1/2}$\,Myr
\citep{2012ApJ...750...13C,2015ApJ...808...48B}. 
Due to the short timescale of evolution, the gravitational collapse could barely affect the B-$\rho$ relation in the simulation
The size of these molecular clouds is around 4.6$^3$ pc (256$^3$ pixels), with one-pixel size of approximately 0.018 pc \citep{2015ApJ...808...48B}.

In this Enzo simulation, the thermal energy is located at the low state (${\cal M}_s$ = 9).
Their timescale of evolution stage is 0.6 $t_{\rm ff}$, around  0.76 Myr, which the timescale starts from when interstellar media collapse.
Due to the short evolutionary time ($\sim$ 3.85 $\times$ 10$^{-21}$ g\,cm$^{-3}$, or 821 cm$^{-3}$), the gravitational energy less affects the system in simulation.
With the similar mean density $\rho_0$ in these simulations, the different $\beta_0$ ($\sim$0.2, 2, 20) present the different magnetic field states: strong B-field, stronger B-field (weaker than strong B-field but stronger than weak B-field), and weak B-field.
The Alfvén Mach number ${\cal M}_{\rm A}$ in various simulations can be calculated by the  energy ratio between the total magnetic energy $E_B$ and kinetic energy $E_k$ \citep{2024ApJ...976..209Z}:
\begin{equation}
    {\cal M}_{\rm}=\sqrt{\frac{E_{\rm k}}{E_{\rm B}}} = \frac{\frac{1}{2}\rho \sigma_v^2}{B^2/8\pi} = \sqrt{4\pi\rho}\frac{\sigma_v}{B}
\end{equation}
where the ${\cal M}_{\rm A}$ in strong, medium, and weak magnetic field states can be estimated as 0.31, 0.49, and 1.29, respectively, when the simulation has evolved 0.6 $t_{\rm ff}$.

\subsection{Validity of the Physical Assumption: Magnetic Tension vs. Pressure Force}\label{Ap.B}

We employ the $\beta_0=20$ simulation ($t \approx 0.76$ Myr) to validate the physical assumptions of the Curvature Mapping Method, specifically comparing the roles of magnetic tension ($\boldsymbol{f}_t$) and magnetic pressure force ($\boldsymbol{f}_p$) within the total Lorentz force ($\boldsymbol{f}_L$). 
We first evaluate the global vector similarity in 3D space:
\begin{equation}
    D_{\rm similarity} =\frac{|\boldsymbol{v}_1-\boldsymbol{v}_2|}{|\boldsymbol{v}_1+\boldsymbol{v}2|} \approx 0.5\,\theta_{\rm offset} \end{equation}
where the $\vec{v_1}$ is the vector of Lorentz force and $\vec{v_2}$ could be magnetic tension or pressure force.
Using the similarity metric $D_{\rm similarity} \approx 0.5\theta_{\rm offset}$, we find that both tension and pressure force statistically resemble the Lorentz force structure (see Fig.\,\ref{fig2f}). 
After correcting for projection effects using the probability density function (see Sect.\,\ref{ap3Dto2D}), the mean projected offset angle between magnetic tension and the total Lorentz force is approximately $11^\circ$, while the offset for magnetic pressure force is around $19^\circ$ (Fig.\,\ref{fig2f}). 
This indicates that, on a global scale, for initial condition similar to that of the molecular cloud, the magnetic pressure force and tension align such that the direction of either, e.g. the magnetic tension, can be used to estimate the direction of the {Lorentz} force

\begin{figure*}[h]
    \centering
    \includegraphics[width=16cm]{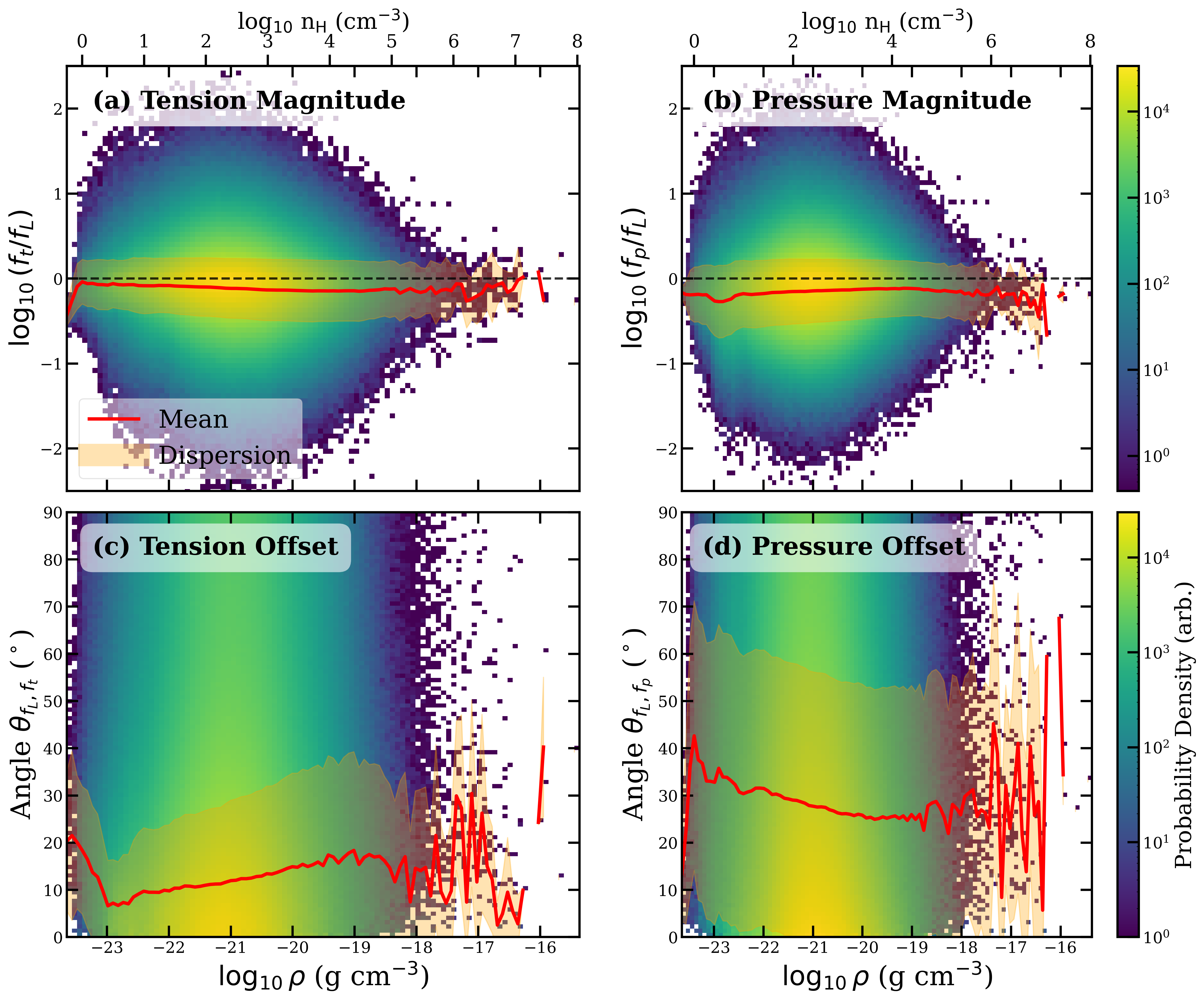}
    \caption{{\bf Statistical comparison of magnetic tension ($f_t$) and magnetic pressure force ($f_p$) relative to the total Lorentz force ($f_L$) in MHD simulations.}
    (a) and (b): Distributions of the magnitude ratios $\log_{10}(f_t/f_L)$ and $\log_{10}(f_p/f_L)$ as a function of gas density. 
    The color scale probability density (number count of pixels). 
    Darker/Yellower colors indicate regions with a higher concentration of pixels.
    (c) and (d): Distributions of the angular offset between the component vectors and the total Lorentz force vector ($\theta_{f_L, f_t}$ and $\theta_{f_L, f_p}$). In all panels, the red solid lines represent the mean value in each density bin, and the orange shaded regions indicate the $1\sigma$ dispersion. While the magnitudes of tension and pressure are comparable (top panels), the magnetic tension vector (traced by curvature) shows significantly better alignment with the total Lorentz force (mean offset $\lesssim 20^\circ$) compared to the magnetic pressure (mean offset $\sim 40^\circ$) across the dense gas regime. }
    \label{figftvsfL}
\end{figure*}

\begin{figure*}[h]
    \centering
    \includegraphics[width = 14cm]{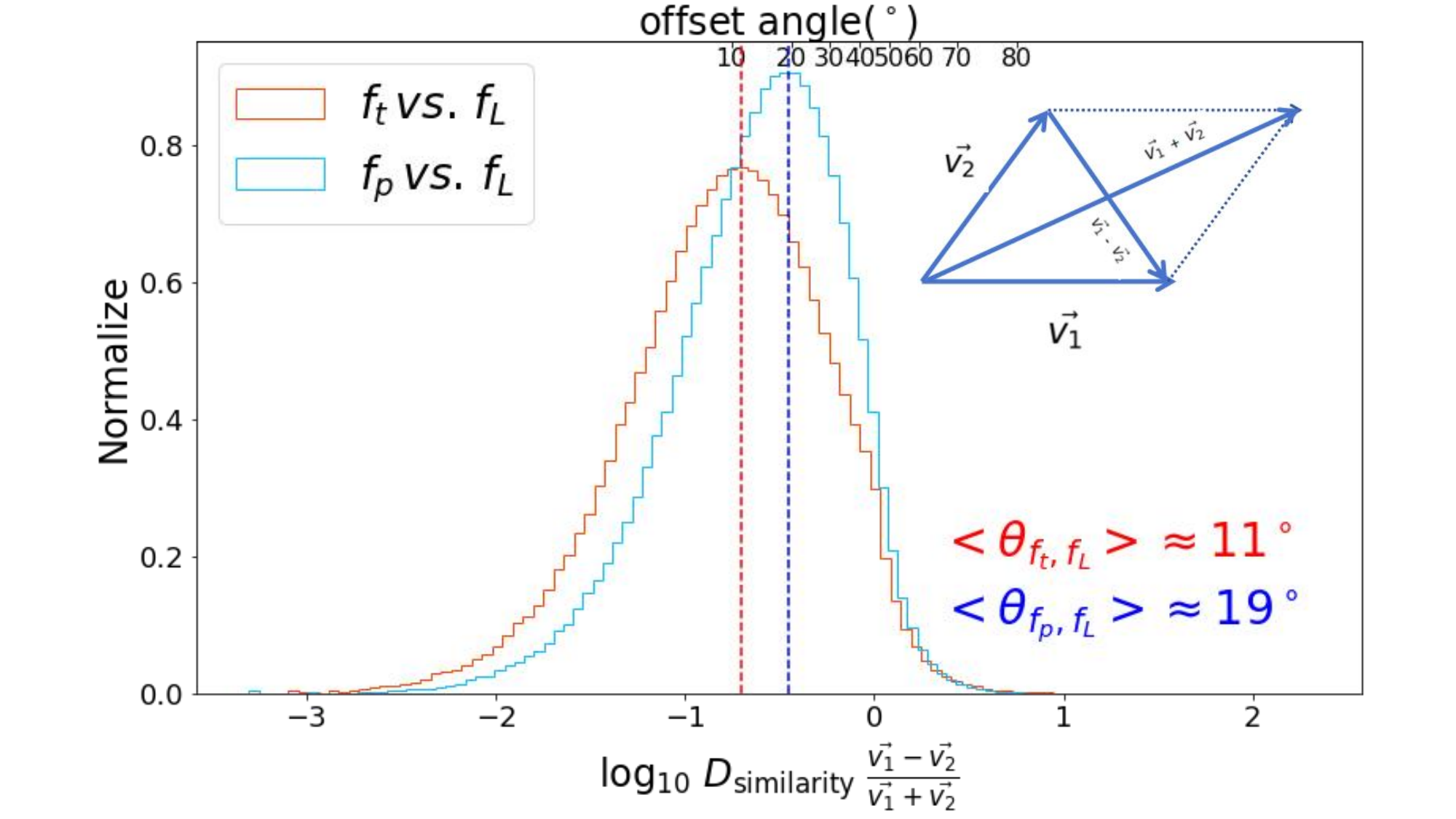}
    \caption{{\bf Distribution of similarity and offset angle between Lorentz force and magnetic tension, magnetic pressure force.}
    By projecting from 3D space to the 2D plane, magnetic tension, or magnetic pressure force, $\vec{V_1}$ is close to Lorentz force $\vec{V_2}$, whose offset angle {\mk between magnetic tension/pressure and Lorentz force} is around 11$^\circ$, {\mk 19$^\circ$, respectively} .
    The canton in the top right corner displays the similarity of two vectors.
    {\mk The Lorentz force is similar to the magnetic tension and pressure force in order.}}
    \label{fig2f}
\end{figure*}

To rigorously assess the physical validity of our method, we investigated the density dependence of these correlations using the full 3D vector fields. Figure\,\ref{figftvsfL} presents the statistical comparison of magnitude ratios and angular offsets as a function of gas density.
Regarding magnitude, both magnetic tension and pressure force remain comparable to the total Lorentz force across the entire density range. 
This confirms that the magnetic pressure force contribution is non-negligible in strength, justifying the introduction of the weighting factor $\Omega$ (Eq.,\ref{eqcurvforce}) to recover the total force magnitude.
The distribution of angular offsets are both below 30$^\circ$ in the dense gas regime ($\log n_{\rm H} \gtrsim 3$). 
Note that produce the distribution of angular offset,  we have taken corrected for the geometric projection effect (see Ap.\,\ref{ap3Dto2D}) to reveal the true physical alignment.
The magnetic tension vector maintains a tight alignment with the total Lorentz force, characterized by a corrected mean offset of $\sim 15^\circ\pm5^\circ$ (Fig.\,\ref{figftvsfL}c). 
In contrast, the magnetic pressure force exhibits significantly larger misalignment ($\sim 30^\circ$) and broader dispersion in the same regime (Fig.\,\ref{figftvsfL}d). 

The choice of curvature (the geometric indicator of magnetic tension) as a tracer is driven not only by this alignment advantage but also by observational feasibility. 
The magnetic pressure force depends on the spatial gradient of the field strength ($\nabla B^2$), which is notoriously difficult to constrain in observations \citep{1951PhRv...81..890D,1953ApJ...118..113C,2021A&A...647A.186S,2022ApJ...925...30L}. 
In contrast, magnetic tension is geometrically determined by the field curvature ($\boldsymbol{\kappa}$), a vector quantity that can be robustly derived from the magnetic field morphology traced by dust polarization (see Sect.\,\ref{method}).
Therefore, combining the intrinsic directional alignment with observational accessibility, magnetic curvature stands out as the optimal proxy for mapping the Lorentz force orientation in molecular clouds.
We also verified the analysis by projecting the simulation data along the x-z and y-z planes. The statistical results remain consistent across these orthogonal projections, confirming the robustness of the method.

\begin{figure*}
    \centering
    \includegraphics[width=0.95\linewidth]{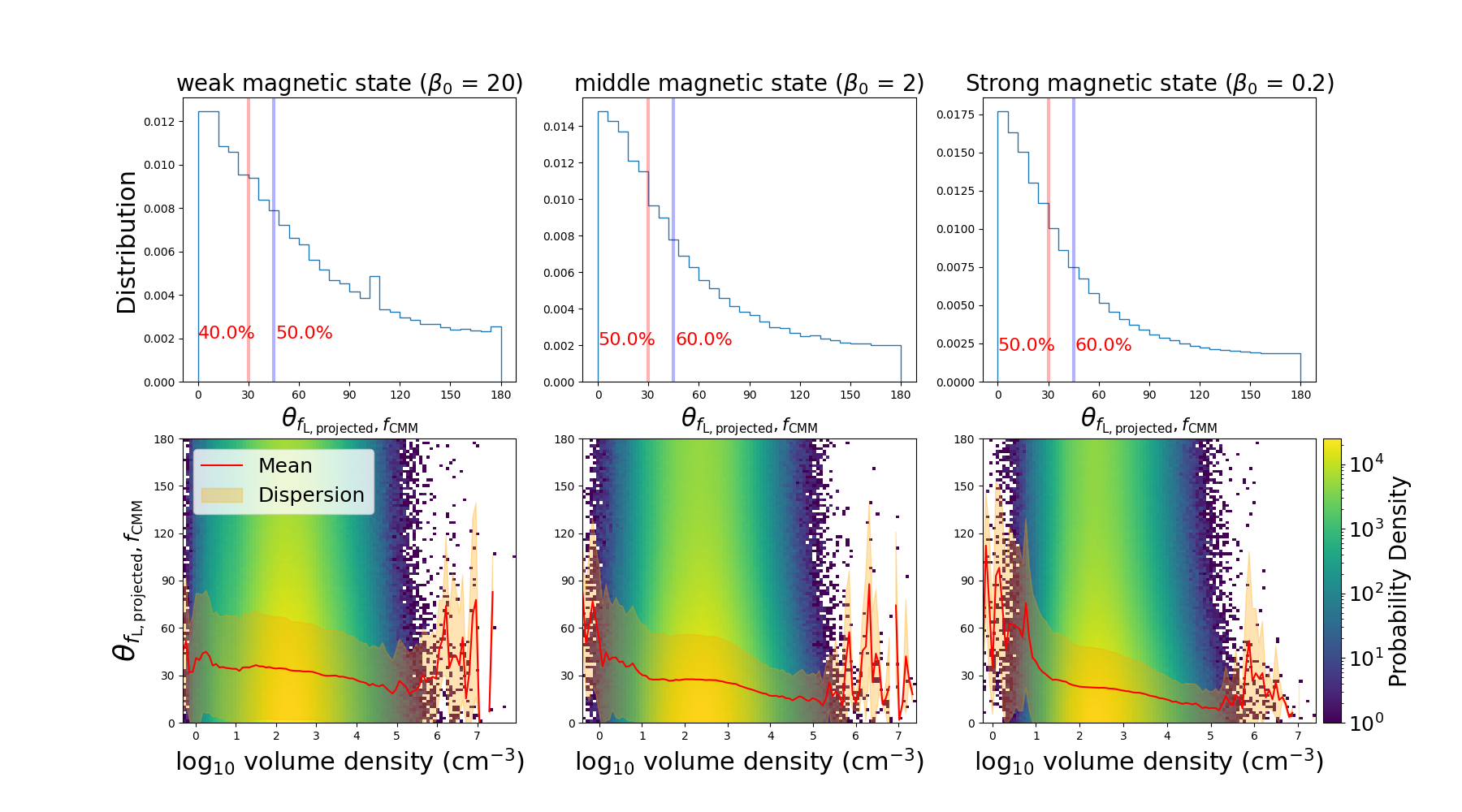}
    \caption{{\bf Distribution of offset angle between {projected Lorentz force (estimated)} and Projected Lorentz force in MHD simulations.}
    The three columns present the distribution of offset angle between {projected Lorentz force (estimated)} and Projected Lorentz force in weak, middle, and strong magnetic field states, respectively.
    The top panels show the offset angle distribution between {projected Lorentz force (estimated)} and Projected Lorentz force in three simulations.
    The percentages of offset angle $<$ 30$^\circ$ (parallel aligned) are around 40$\%$, 50$\%$ pixels, and 50$\%$ in weak, middle, and strong magnetic field states, respectively.
    That of offset angle $<$ 45$^\circ$ (close to parallel) are around 50$\%$, 60$\%$, and 60$\%$ in weak, middle, and strong magnetic field states, respectively.
    The bottom panels show the distribution between volume density and offset angle, where the red lines display the mean trend at each density bins and the orange cover range shows the dispersion.}
    
    \label{figE1}
\end{figure*}

\begin{figure*}[h]
    \centering
    \includegraphics[width=\linewidth]{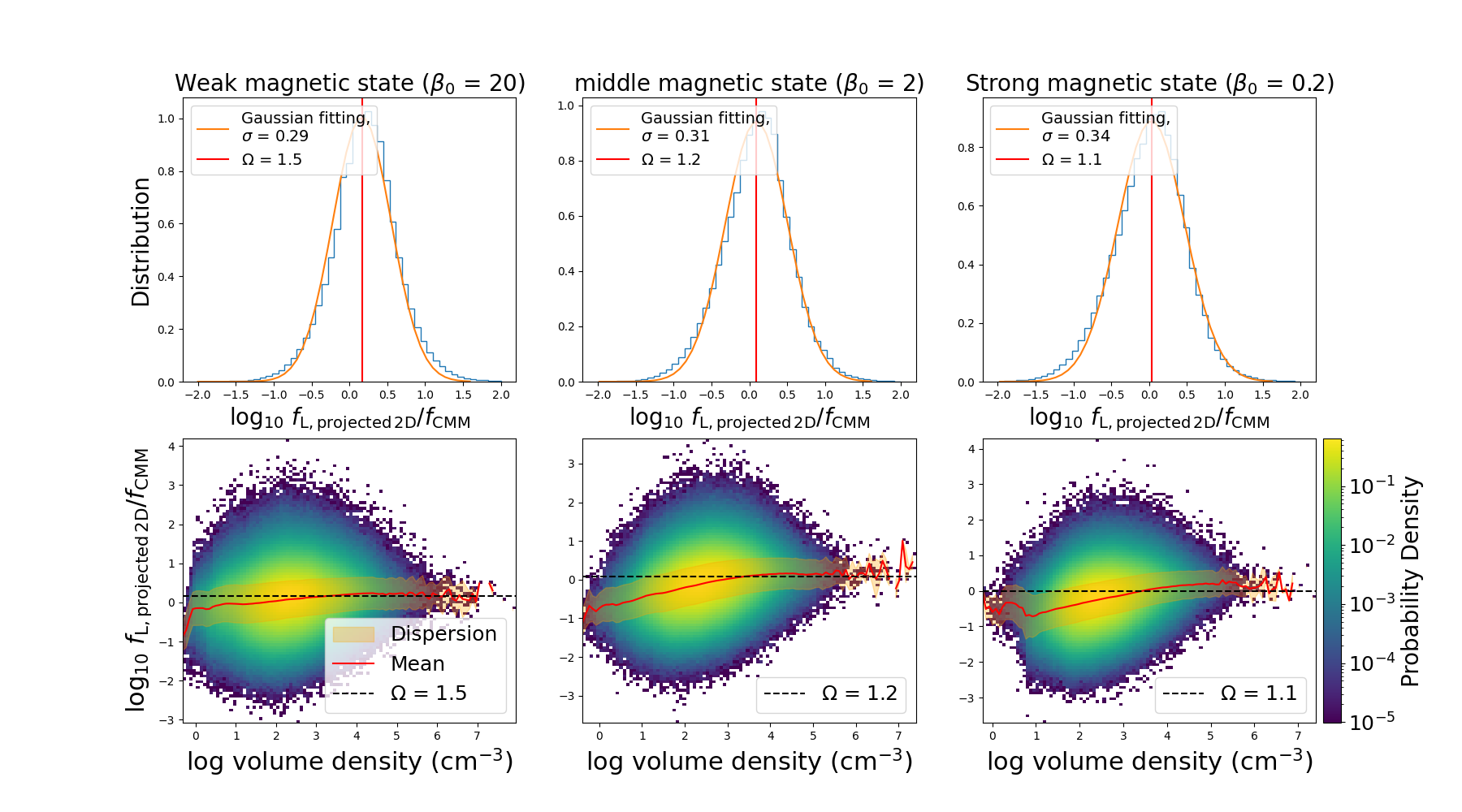}
    \caption{{\bf Distribution of modified factors of curvature mapping method in various magnetic field states.}
    The ratio between the projected Lorentz force and the {projected Lorentz force (estimated)} presents the modified factor $\Omega$ of curvature mapping method.
    In three simulations (weak, middle, and strong magnetic state in left, middle, and right panel), the distribution of modified factor forms norm distributions, where the mean values are 1.5, 1.2, and 1.1 with the standard deviation around 0.3 at the log scale. 
    The top panels show the 1D histogram of this ratio and the bottom panel show the changing of this ratio with density growth.}
    \label{fig3}
\end{figure*}

\subsection{Accuracy of Curvature Mapping Method in Determining Lorentz Force}\label{sect3.3}

To verify the accuracy of the Curvature Mapping Method in tracing the POS projected component of the Lorentz force, we compare the force derived from our method (referred to as Sect.\,\ref{method}) with the true projected component of the 3D Lorentz force calculated directly from the simulation.

The components of the true 3D Lorentz force are given by:
\begin{equation}
\boldsymbol{f}_{L} = \frac{1}{\mu_0}(\nabla \times \boldsymbol{B}) \times \boldsymbol{B}\, ,
\end{equation}
The projected Lorentz force in the $x$-$y$ plane is its projection:
\begin{equation}
\boldsymbol{f}_{L, \rm projected} = f_{L,x}\hat{\boldsymbol{i}} + f_{L,y}\hat{\boldsymbol{j}} \,,
\end{equation}
The {CMM-derived projected Lorentz force} is computed {following} the observational procedure, utilizing only the unit vector of the magnetic field in the $x$-$y$ plane and the total magnetic field strength:
\begin{equation}
\boldsymbol{f}_{\rm CMM} = \frac{B{\rm est}^2}{\mu_0} (\boldsymbol{b} \cdot \nabla) \boldsymbol{b} = \frac{B_{\rm est}^2}{\mu_0} \boldsymbol{\kappa}{\rm POS},
\end{equation}
where $\boldsymbol{b} = b_x \hat{\boldsymbol{i}} + b_y \hat{\boldsymbol{j}}$ is the unit vector of the magnetic field in the $x$-$y$ plane ($b_x = B_x/\sqrt{B_x^2+B_y^2}$), and $B_{\text{est}} = \frac{4}{\pi} B_{\text{POS}}$\citep{2004ApJ...600..279C} ($B_{\text{POS}} = \sqrt{B_x^2 + B_y^2}$ is the projected field strength).
The $x$ and $y$ components of the {CMM-derived projected Lorentz force} are calculated as:
\begin{equation}
f_{\rm x,CMM} = \frac{B_{\rm est}^2}{\mu_0} \left( b_x \frac{\partial b_x}{\partial x} + b_y \frac{\partial b_x}{\partial y} \right)\,,
\end{equation}
\begin{equation}
f_{\rm y,CMM} = \frac{B_{\rm tot}^2}{\mu_0} \left( b_x \frac{\partial b_y}{\partial x} + b_y \frac{\partial b_y}{\partial y} \right)\,,
\end{equation}
We analyzed 2D slices along the $z$-axis from the $256^3$ MHD simulations to simulate random observational perspectives and evaluate the method's performance.
We also verified the analysis by projecting the simulation data along the x-z and y-z planes. The statistical results remain consistent across these orthogonal projections, confirming the robustness of the method.

\subsubsection{Comparison of Directions}

We first evaluate the directional accuracy by calculating the offset angle $\theta$ between the {estimated projected Lorentz force} ($\boldsymbol{f}_{\rm CMM}$) and the projected Lorentz force ($\boldsymbol{f}_{L, \rm projected}$) in the image plane.

Figure\,\ref{figE1} displays the global distribution of offset angles for weak, middle, and strong magnetic field cases.
Globally, approximately 50$\%$-60$\%$ of the pixels exhibit a "close-to-parallel" alignment ($\theta \lesssim 45^\circ$).
However, a critical trend is revealed when analyzing the error as a function of gas density (Fig.\,\ref{figE1}, bottom panels). 
In diffuse regions (volume density $<$ 10$^2$ cm$^{-3}$), turbulence randomizes the magnetic structure, leading to large misalignments ($<\theta_{f_{\rm L,projection}, f_{\rm CMM}} \gtrsim$30$^\circ$. 
In contrast, in high-density regions ($\log n_{\rm H} \gtrsim 3$), the alignment significantly improves, with the mean offset angle dropping to $\sim 10^\circ$--$25^\circ$.
This density-dependent performance confirms that while the method may be limited in the diffuse ISM, it effectively and accurately traces the orientation of the Lorentz force in the dense molecular structures relevant to star formation. 

\subsubsection{Strength Modification and the $\Omega$ Factor}\label{sec3.3.2}

Since the Curvature Mapping Method uses a geometric approximation for the force, a modification factor $\Omega$ is introduced to account for the magnetic pressure force contribution and projection effects:
\begin{equation}
\Omega = \frac{|\boldsymbol{f}_{\rm L, projected}|}{|\boldsymbol{f}_{\rm CMM}|}\,,
\end{equation}
As shown in Fig.\,\ref{fig3}, the ratio $\Omega$ follows a log-normal distribution with a dispersion of $\sim 0.3$ dex. Gaussian fitting yields mean values of $\Omega \approx 1.1$, $1.2$, and $1.5$ for magnetic, magnetic-kinetic, {kinetic} region $1.1$, $1.2$, and $1.5$ for strong, middle and weak magnetic states, respectively.
Crucially, the bottom panels of Fig.\,\ref{fig3} demonstrate that this ratio remains remarkably stable across the high-density regime. This stability justifies the use of a constant $\Omega$ (ranging from 1.1 to 1.5) to recover the magnitude of the projected Lorentz force in dense cores.

In summary, the projected Lorentz force can be reliably mapped as: 
\begin{equation} 
    \boldsymbol{f}_{\rm L, projected} \approx \Omega \frac{B_{\rm tot}^2}{\mu_0} \boldsymbol{\kappa}_{\rm POS} \,,
\end{equation} 
where $\boldsymbol{\kappa}_{\rm POS}$ is the curvature vector derived from polarization maps. Given the method's proven accuracy in dense gas, we proceed to apply it to the OMC-1 region, which features a classical 2D "hourglass" magnetic field structure \citep{2017ApJ...846..122P}.

\begin{figure*}[h]
    \centering
    \includegraphics[width=0.9\linewidth]{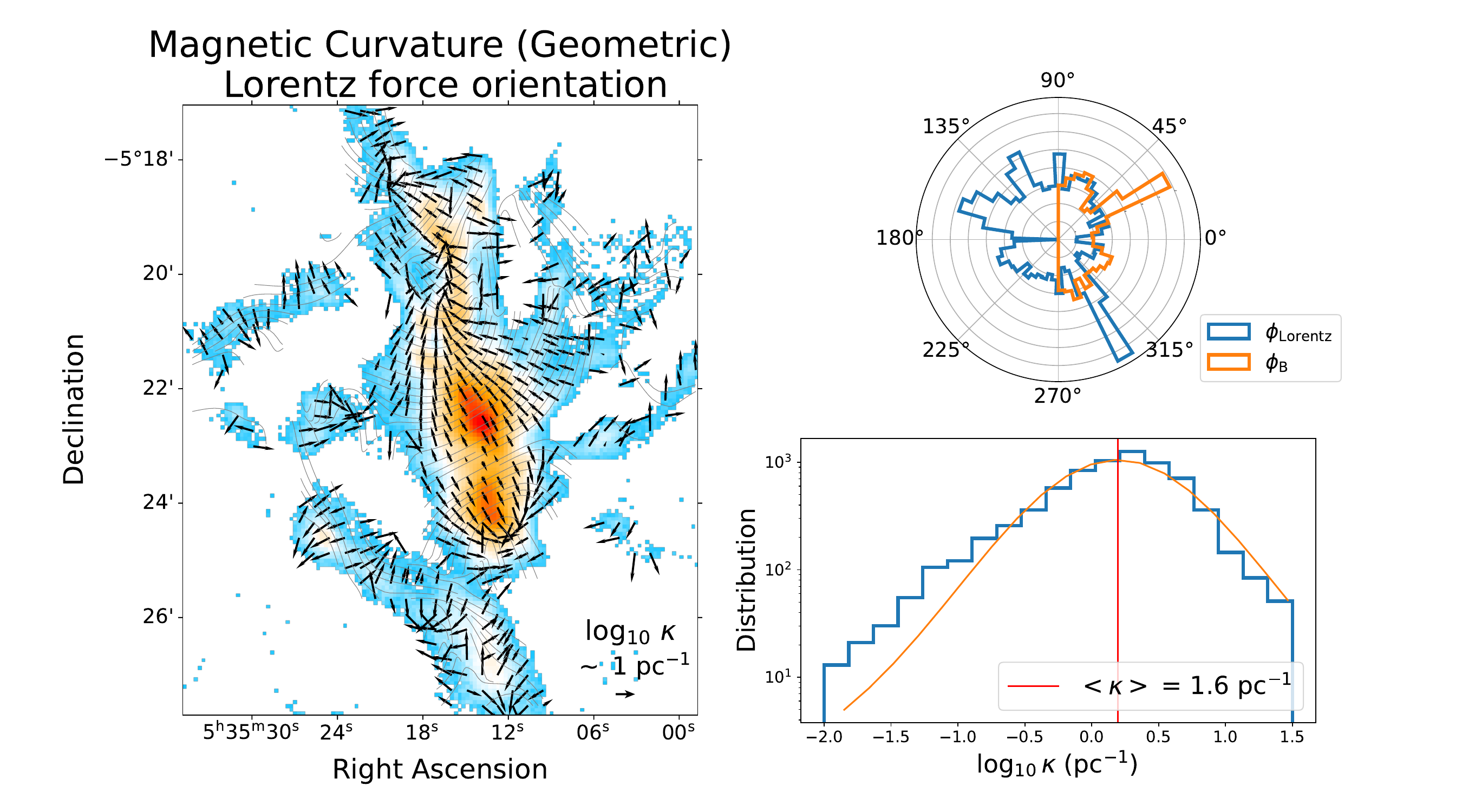}
    \caption{{\bf Distribution of magnetic curvature and Lorentz force orientations.}
    The left panel displays the magnetic curvature map (black arrows), where the arrow orientations represent the orientation of the Lorentz force. The background illustrates the 850$\mu$m continuum emission, while the streamlines depict the magnetic field structure derived from 850$\mu$m dust polarization observations.
    {\mk The contours show the column density as 10$^{23.4}$ cm$^{-2}$ and the triangle markers display the position of clumps Orion-KL and Orion South.}
    The top-right panel shows the orientation distributions of the Lorentz force (blue) and the magnetic field (orange). 
    The Lorentz force orientation peaks at angles perpendicular to those of the magnetic field, indicating the expected orthogonality between the two.
    The bottom-right panel presents the distribution of magnetic curvature magnitudes. 
    The orange line represents a Gaussian fit to the distribution, and the red line indicates the average curvature value ($<\kappa>\sim 1.6{\rm pc}^{-1}$).}
    \label{figcurvature}
\end{figure*}

\begin{figure*}[h]
    \centering
    \includegraphics[width=\linewidth]{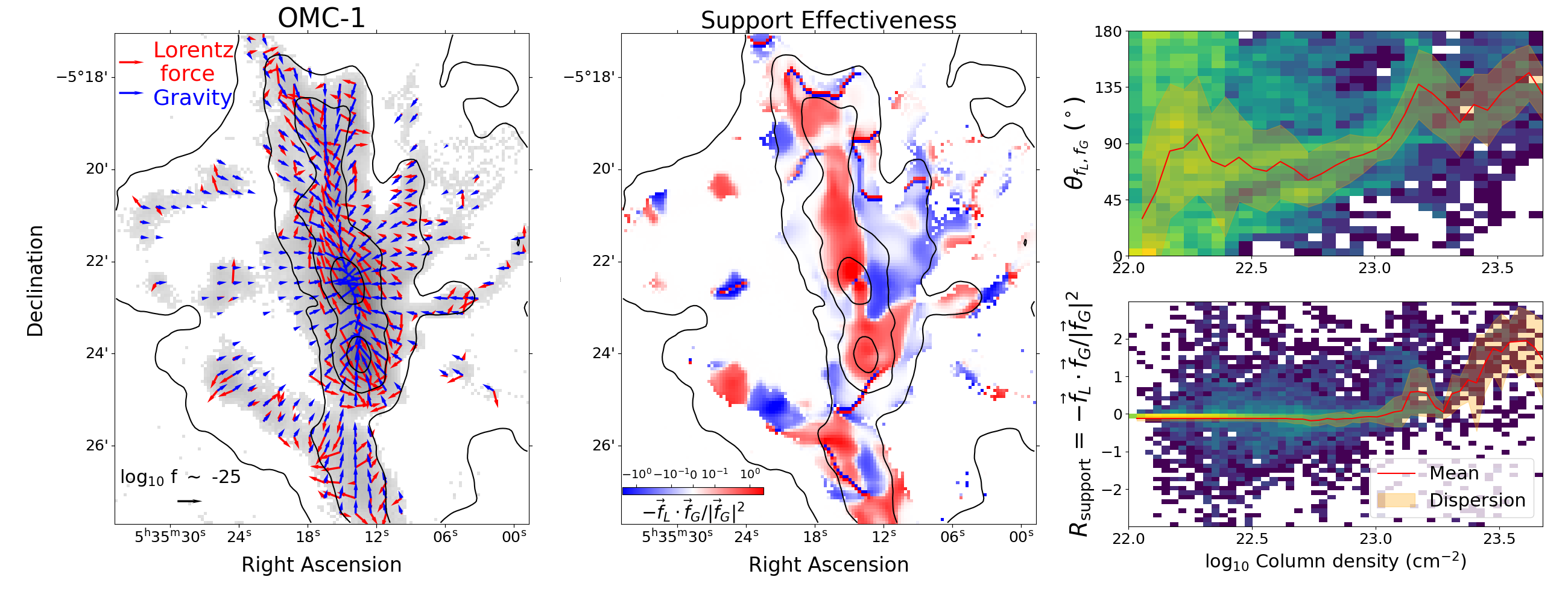}
    \caption{{\bf Interplay between Lorentz force and Gravity.}
    The left panel displays the vector map of the Lorentz force (red arrows) and gravitational force (blue arrows).
    The background illustrates the 850$\mu$m continuum emission, while the streamlines depict the magnetic field structure derived from 850$\mu$m dust polarization observations. 
    Black contours show the H$_2$ column density from Herschel from 10$^{22}$ to 10$^{23.5}$ cm$^{-2}$ with step as 10$^{0.5}$. 
    The bottom-right panel presents the 2D histogram between alignment angle $\theta_{f_{\rm, L}, f_{\rm G}}$ and H$_2$ column density.
    The top-right pane presents the distribution of presents the 2D histogram between magnetic support ratio $\mathcal{R}_{\rm support}$ and H$_2$ column density..
    The red line represents mean trend and orange cover regions show the dispersion (1$\sigma$).
    The middle panel present the spatial distribution of magnetic support ratio $\mathcal{R}_{\rm support}$.
    }
    \label{figmap}
\end{figure*}

\section{Mapping Lorentz Force in Orion A}

Orion A is a well-known massive star-forming region located approximately 400\,pc away \citep{2007A&A...474..515M}. 
Within this molecular cloud lies the OMC-1 region, featuring a distinctive pinched magnetic field structure (see Fig.\,\ref{figmap}), which is a hallmark of hourglass magnetic field configurations.

By applying the Curvature Mapping Method to magnetic field observations of OMC-1 \citep{2008A&A...487..247F, 2017ApJ...846..122P}, we constructed a Lorentz force map as vector fields within the molecular cloud. 
The vector form of the projected Lorentz force is determined using the Curvature Mapping Method, which relies on both scalar term {magnetic field strength} and vector terms {magnetic curvature}:
\begin{equation} 
f_{\rm L} = \frac{\Omega}{\mu_0}{B_{\rm tot}^2}\boldsymbol{\kappa}\,, 
\end{equation}
where the orientation of the projected Lorentz force is aligned with the direction of the magnetic curvature, and its strength is predominantly influenced by the magnitude of the magnetic field strength and the magnetic curvature. 
We calculate the local curvature $\kappa$ purely from the geometric morphology of the magnetic field lines. Specifically, we derive the orientation angle $\phi_B = 0.5 \arctan(U/Q)$ from the observed Stokes $Q$ and $U$ parameters. 
We then apply Equations\,\ref{eqkx}, \ref{eqky}, and \ref{eqkxy} to compute the spatial derivatives of $\phi_B$, yielding the projected curvature vector. 
This calculation is performed pixel-by-pixel for all regions satisfying our signal-to-noise criteria ($I/\sigma_I > 3$ and $P/\sigma_P > 3$), without any prior assumptions about the magnetic field strength or Lorentz force

With reliable magnetic field structures from observations, the magnetic curvature can be readily derived from observations (see Sect.\,\ref{method}).
However, estimating magnetic field strength remains a significant observational challenge. 
To address this limitation, we propose two approaches for studying the Lorentz force from observations:
\begin{itemize} 
\item Determination of the map of Lorentz force structure (orientation) {\mk and the size of magnetic curvature}. 
\item Full vector mapping of the Lorentz force, encompassing both orientation and strength. 
\end{itemize}

\subsection{Spatial Structure of Lorentz Force}\label{FLmorp}

The directions and structure of Lorentz force can be traced by the orientation of magnetic curvature derived by magnetic field morphology.
As Fig.\,\ref{figcurvature} shows, the Lorentz force exhibits a structured, bipolar geometry in OMC-1 with an hourglass-shaped magnetic field.
The observed Lorentz force distribution highlights localized regions of high curvature in the magnetic field, where the force vectors often diverge in opposing directions, forming a bipolar pattern. 
This structure suggests that the Lorentz force plays a critical role in regulating gas dynamics and counteracting gravitational collapse in these regions.

In the context of star formation, this interplay between Lorentz force and gravity is crucial. 
The Lorentz force in the hourglass-shaped magnetic field acts as a counterbalance to gravity, slowing down the collapse of gas and dust in molecular clouds \citep{1993ApJ...417..220G,2003ApJ...599..351A,2014ApJ...794L..18Q}. 
This magnetic support is particularly important in regions of high curvature, where the magnetic field's resistance to distortion is strongest. 
By buffering against gravitational collapse, the Lorentz force helps regulate the rate of star formation, preventing rapid and unregulated fragmentation of the cloud \citep{2017A&A...602L...2H}.

The orientation of the force is theoretically perpendicular to its potential field. 
As Fig.\,\ref{figcurvature} shows, the orientation of Lorentz force is perpendicular to the magnetic field both POS plane morphology and statistics polar space of orientations.
By ensuring the magnetic field lines and the Lorentz force vectors are consistently orthogonal, we can be more confident in the robustness of the curvature mapping method used to derive the force structures in these regions.

\subsection{{Lorentz Force vs. Gravity in OMC-1}}
\subsubsection{Mapping Lorentz force in OMC-1}

If a reliable magnetic field strength distribution is available, the Lorentz force can be mapped in vector form (see Fig.\,\ref{figmap}). 
We employ the ST method \citep{2021A&A...647A.186S,2021A&A...656A.118S} to derive the scalar magnetic field strength map based on gas density, non-thermal velocity dispersion, and the angular dispersion of the magnetic field orientation (input parameter details are shown in Ap.\,\ref{Ap.estB}).

To derive the magnitude of Lorentz force, an independent measurement of the magnetic field strength ($B$) is required as an input. Therefore, 
we utilize the Structure Function (ST) method to derive the mean magnetic field strength (see Fig.\,\ref{figD1}).
By combining the spatial distribution of magnetic field strength ($B_{\rm tot}$) and the vector map of magnetic curvature ($\boldsymbol{\kappa}$; see Fig.\,\ref{figcurvature}), the Lorentz force vectors are calculated as:
\begin{equation}
\boldsymbol{f}_{\rm L} = \frac{\Omega}{\mu_0}B_{\rm tot}^2\boldsymbol{\kappa}\,,
\end{equation}
where the correction factor $\Omega$ is taken as 1.5, consistent with the value derived for the weak-field case (see Sect.,\ref{sec3.3.2}, \citealt{2024ApJ...977L..31W,2024ApJ...976..209Z}).
The resulting Lorentz force vector field is displayed in Fig.\,\ref{figmap} (red arrows).

\subsubsection{Spatially Resolved Gravity Vectors}\label{sec:gravity}

To compare {the} Lorentz force and gravity, we calculated the gravitational potential ($\Phi$) directly from the column density map and its gradient, gravitational acceleration.
Following the 2D gravitational potential measurement \citep{2023MNRAS.526L..20H}, we solved the Poisson equation in Fourier space assuming a simplified geometry appropriate for the projected data:
\begin{equation}
\Phi_{k} = - \frac{2\pi G \Sigma_k}{|\mathbf{k}| (1 + |\mathbf{k}| H_{\text{eff}})}\,,
\end{equation}
where $\Sigma_k$ is the Fourier transform of the mass surface density ($\Sigma = \mu m_H N_{H_2}$), $\mathbf{k}$ is the wave vector, and $H_{\text{eff}}$ represents an effective half-thickness (0.2 pc, filament width). 
This approach allows us to map the gravitational acceleration on POS:
\begin{equation}
    \boldsymbol{a}_{\rm G} = - \nabla \Phi\,,
\end{equation}
The local volume density of the gravitational force is obtained by coupling the acceleration with the volume density ($\rho$, derived in Ap.\ref{Ap.estB}):
\begin{equation}
    f_{\rm G} = \rho a_{\rm G} = \mu {\rm m_H} {\rm n_{\rm H}} \cdot a_{\rm G}\,,
\end{equation}
where $n_{\rm H}$ is volume density, detail shown in Fig.\,\ref{figD1}.
The vector of gravitational force is plotted as blue vector in Fig.\,\ref{figmap}.

\subsubsection{Magnetic Support Effectiveness} 

The spatially resolved maps of $\boldsymbol{f}_{\rm L}$ and $\boldsymbol{f}_{\rm G}$ allow for a direct, pixel-by-pixel evaluation of the interplay between magnetic fields and gravity.
In contrast to the polarization orientation with limited dynamic range ($0-\pi$), both the Lorentz force and gravitational force are polar vectors defined over the full $0-2\pi$ range. 
This enables us to determine whether the magnetic force is supporting against ($\theta_{f_{\rm L}, f_{\rm G}}$ close to 180$^\circ$) or assisting the gravitational collapse ($\theta_{f_{\rm L}, f_{\rm G}}$ close to 0$^\circ$).

Fig.\,\ref{figmap} presents the distribution of the alignment angle, $\theta(\boldsymbol{f}_{\rm L}, \boldsymbol{f}_{\rm G})$, as a function of column density. 
A clear bimodal behavior is observed.
In the diffuse envelope ($\log N_{\rm H_2} \lesssim 23$ cm$^{-2}$), the distribution is broad with a mean angle near $90^\circ$, indicating that the magnetic force hardly affect gravity.
In the dense filament ridge ($\log N_{\rm H_2} \gtrsim 23.0$ cm$^{-2}$), the alignment transitions sharply to an anti-parallel configuration, with the angle approaching $180^\circ$. 
The magnetic field is resisting gravitational collapse in the dense gas.

To quantify this interplay between gravity and magnetic force, we calculate the support effectiveness ratio, defined as the projection of the Lorentz force onto the direction opposing gravity, normalized by the gravitational force magnitude:
\begin{equation}
\mathcal{R}{\rm support} = \frac{-\boldsymbol{f}_{\rm L} \cdot \boldsymbol{f}_{\rm G}}{|\boldsymbol{f}_{\rm G}|^2}\,,
\end{equation}
As shown in the Fig.\,\ref{figmap}, the support ratio $\mathcal{R}_{\rm support}$is close to unity in the high-density ridge, where the projected component of the Lorentz force is comparable to the magnitude of gravity
Considering the typical factor-of-two uncertainty in magnetic field strength estimation (which translates to a factor of $\sim 4$ uncertainty in force magnitude), the mean ratio in the ridge reaches $\mathcal{R} \sim 0.5-1.0$.
Spatially, the support effectiveness map (Fig.\,\ref{figmap}, middle panel) reveals that the "backbone" of the Orion A filament (shown in red) experiences strong magnetic support ($\mathcal{R} > 0$), while the surrounding regions (shown in blue) are dominated by gravitational infall.
This suggests that while the region may be collapsing \citep{2017A&A...602L...2H}, the magnetic force provides substantial support that likely regulates the fragmentation and collapse rate of the filament.

\subsection{{Constraints on Magnetic Field Strength}}\label{sec4.2}

Direct measurement of magnetic field strength in the interstellar medium (ISM) remains challenging. 
However, the Curvature Mapping Method provides a way to estimate the \textit{equilibrium magnetic field strength} ($B_{\rm eq}$) required to balance the local self-gravity.
By equating the magnitude of the Lorentz force density ($f_L$) to the gravitational force density ($f_{\rm G}$), the magnetic field strength can be constrained as:
\begin{equation}
    f_L = \frac{\Omega}{\mu_0}B_{\rm eq}^2 \kappa \approx f_{\rm G} \,,
\end{equation}
which yields:
\begin{equation}
    B_{\rm eq} \approx \sqrt{\frac{\mu_0 f_{\rm G}}{\Omega \kappa}} \,,
\end{equation}
This $B_{\rm eq}$ represents the theoretical magnetic field strength needed to support the cloud against gravitational collapse. For collapsing regions, this value serves as a physical reference for the field strength required to maintain hydrostatic equilibrium.

We apply this estimation to the OMC-1 region. 
Using the spatially resolved gravitational force map derived in Sect.\,\ref{sec:gravity} (see Fig.\,\ref{figmap}), the average gravitational force density in the dense ridge is approximately $<f_{\rm G}> \sim 2 \times 10^{-23}$ dyne cm$^{-3}$. 
Combining this with the average magnetic curvature derived from the CMM ($<\kappa> \approx 1.6$\,pc$^{-1}$, see Fig.\,\ref{figcurvature}) and adopting $\Omega = 1.5$, Eq.\,\ref{eqcurvforce} yields a required field strength of $B_{\rm eq} \approx 0.5$\,mG.

This estimated equilibrium value is remarkably consistent with independent measurements, including Zeeman splitting ($\sim 0.7$ mG, \citealt{2010ApJ...725..466C}) and the Davis-Chandrasekhar-Fermi (DCF) method ($\sim 0.6$ mG; \citealt{2024ApJ...977L..31W}). 
This consistency suggests that the magnetic field in OMC-1 is strong enough to be comparable to gravity, yet the region remains dynamically active.

\section{Conclusion}

We propose a new technique, the Curvature Mapping Method (CMM), to obtain the vector map of the projected Lorentz force in the interstellar medium on the plane of the sky. 
The CMM effectively traces the restoring magnetic forces, provided that the magnetic field structure is shaped by the competition against gravity (or other external forces), rather than being in a force-free configuration.
The method is based on the idea that the Lorentz force is closely related to magnetic tension, which depends on the morphology and strength of the magnetic field:
\begin{equation}
    \boldsymbol{f}_{\rm L} = \frac{\Omega}{\mu_0} B_{\rm tot}^2\cdot \boldsymbol{\kappa}  \,, \nonumber
\end{equation}
where the magnetic field strength $B_{\rm tot}$ is the scalar term, the magnetic curvature $\boldsymbol{\kappa}$ is the vector term, and the modified factor $\Omega$ for the projected effect ranges from [1.1, 1.5]. 
The magnetic curvature, calculated from the magnetic field morphology, is a critical parameter of the Curvature Mapping Method and determines the orientation of the POS Lorentz force.

We verified the accuracy of the method using 3D MHD simulations.
While the method shows moderate accuracy in the diffuse ISM, it demonstrates remarkable reliability in star-forming regions.
Specifically, in high-density gas ($\log n_{\rm H} > 3$ cm$^{-3}$), the derived force direction aligns with the true projected Lorentz force (offset $< 30^\circ$), and the magnitude is recovered within a factor of $\sim 2$.

Applying the Curvature Mapping Method to observations, there are three usage scenarios depending on the input parameters:
\begin{itemize} 
\item Force Orientation: Based on credible magnetic field morphology (e.g., from dust polarization), the CMM can accurately determine the magnetic curvature and the orientation of the POS Lorentz force
\item Field Strength Constraint: Based on the magnetic curvature, the magnetic field strength required to balance gravity can be estimated, providing a theoretical upper limit for collapsing regions.
\item Vector Force Mapping: If a credible magnetic field strength map is available, the full vector map of the Lorentz force can be derived.
\end{itemize}

Applying the CMM to the OMC-1 region, combined with spatially resolved magnetic field strength (derived via the ST method) and self-gravity maps, we performed a pixel-by-pixel analysis of the dynamic interplay between magnetic fields and gravity. Our main findings are:
\begin{itemize} 
\item Vector Mapping: We successfully constructed the first 2D vector map of the Lorentz force in the Orion A filament.
\item Bimodal Behavior: We identified a clear transition in the magnetic force's role. In the diffuse envelope, the magnetic force is statistically uncorrelated with gravity.
\item Magnetic Support: Along the dense filamentary ridge, the Lorentz force vectors are systematically anti-parallel to the gravitational acceleration vectors. The quantitative support ratio is found to be $\mathcal{R}_{\rm support} \sim 0.5 - 1.0$.
\end{itemize}

These results provide direct observational evidence that the magnetic field offers substantial support against gravitational collapse in the dense backbone of OMC-1, likely regulating the star formation rate. With modern polarization instruments (e.g., JCMT, ALMA), the Curvature Mapping Method offers a powerful new tool to visualize and quantify the dynamical role of magnetic fields in the interstellar medium.

\bibliographystyle{aa}
\bibliography{sn-bibliography}

\appendix

\section{{Observational Data}}\label{adata}

The polarized data come from the B-fields In STar-forming Region Observations (BISTRO) survey \citep{2017ApJ...842...66W, 2017ApJ...846..122P}
The resolution of the polarization data is 14$''$ (FHWM $\sim$ 0.03\,pc) measured by Scuba-2+POL-2 at 850$\mu$m.
To calculate the polarized angle, the Stokes parameters Q and U are used in the equation:
\begin{equation}
    \begin{aligned}
        \rm \phi_{p} = 0.5 \times arctan(U,Q) \,,
    \end{aligned}
\end{equation}
where polarized angle $\phi_{p}$ varies from -90$^\circ$ to 90$^\circ$. 
This is the IAU convention. 
One has to use $\phi$=0.5 $\times$ arctan(-U, Q) to convert Planck measurement to this IAU convention \citep{2020A&A...641A...1P}.
The magnetic field (after that, B-field) orientation can be obtained by adding 90$^\circ$ to the polarized angle: $\phi_B$ = $\phi_{\rm p}$ + 90$^\circ$.

The H$_2$ column density map in the molecular cloud, OMC-1, comes from Herschel Gould Belt Survey \footnote{\url{http://www.herschel.fr/cea/gouldbelt/en/Phocea/Vie_des_labos/Ast/ast_visu.php?id_ast=66}}\citep{2010A&A...518L...2P,2010A&A...518L...3G,2013ApJ...763...55R,2013ApJ...777L..33P}, which apply the SED fitting procedure \citep{2015MNRAS.450.4043W} on the Herschel continuum at wavelengths of 70, 160, 250, 350, 500$\mu$m.

\section{{Correcting Projection Effects in Probability Density Distributions}} \label{ap3Dto2D}


The probability density function, p($\theta$), of the angle between two vectors of random distribution is distributed in an N-dimensional space as:
\begin{equation}
    p(\theta) = \frac{\Gamma(\frac{n}{2})}{\Gamma(\frac{n-1}{2})} \frac{{\rm sin}^{n-2}(\theta)}{\sqrt{\pi}}
\end{equation}
where n is the number of dimensions, and $\theta$ is the angle between two vectors of random distribution.


In 3-dimension space, the probability density function of angle p($\theta$) is 
\begin{equation}
    p(\theta) = \frac{1}{2} {\rm sin} \theta
\end{equation}

To remove the projection effect, we use the probability density function p($\theta$) to weight the angle distribution: 
\begin{equation}
   P( \theta_{\rm corrected} )= P(\theta_{\rm origininal}) / p(\theta)
\end{equation}
where $\theta_{\rm original}$ represents the original angle distribution and $\theta_{\rm corrected}$ represents the corrected angle distribution, with the projected effects removed.

\section{{Estimation of Magnetic Field Strength Distribution}}\label{Ap.estB}

\begin{figure*}[h]
    \centering
    \includegraphics[width=0.95\linewidth]{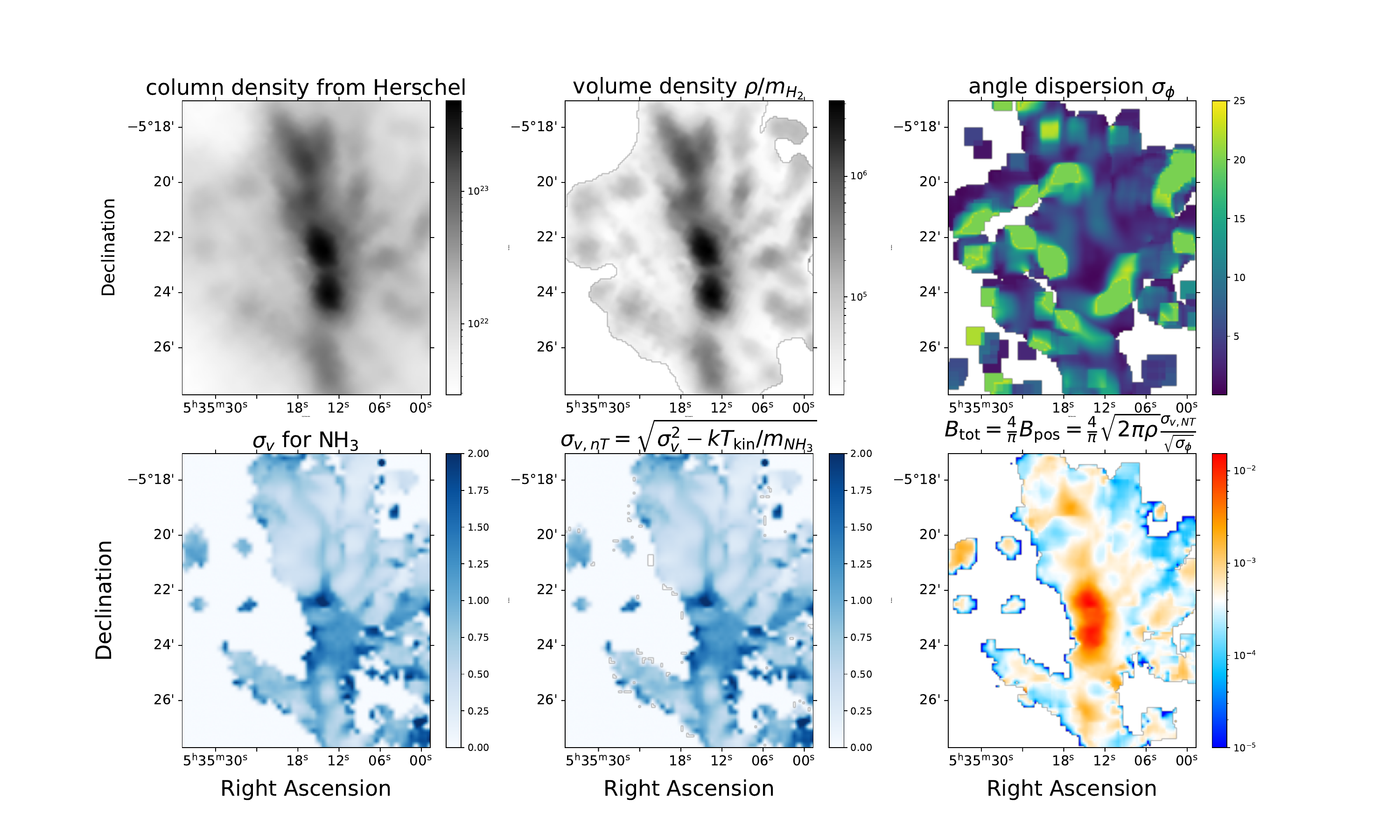}
    \caption{{Estimate the magnetic field strength distribution.}
    We use the ST method to estimate the magnetic field strength.
    These panels present the distribution of column density, volume density, }
    \label{figD1}
\end{figure*}

Accurately measuring the distribution of magnetic field strength on POS is challenging due to limitations in existing methods, such as Zeeman splitting \citep{2010ApJ...725..466C} and the Davis-Chandrasekhar-Fermi (DCF) method \citep{1951PhRv...81..890D,1953ApJ...118..113C,2004ApJ...600..279C}. 
Zeeman splitting hardly maps magnetic field strength effectively which observations usually come from a single point source due to a low signal-to-noise ratio. 
The DCF method ($B_{\rm pos} = 0.5\sqrt{4\pi}\sigma_v/\sigma_\phi$) relies on the assumption of equipartition between turbulent and magnetic energy and turbulence cause the distortion of straight magnetic field line, introducing errors from various observational uncertainties such as density ($\rho$) derived from continuum data\citep{2025arXiv250801130Z}, velocity dispersion ($\sigma_v$), and magnetic field orientation angle dispersion ($\sigma_\phi$) derived from polarization measurements. 
The DCF combined with the Autocorrelation Function (ACF,\citep{2009ApJ...706.1504H,2016ApJ...820...38H}) effectively avoids the overestimation of magnetic field distortion from turbulence but only provides one value of magnetic field strength in single clouds.

To measure the scale of POS Lorentz force, we choose the ST method \citep{2021A&A...647A.186S,2021AA...647A..78A} to estimate the plane distribution of magnetic field strength:
\begin{equation}
    B_{\rm pos} = \sqrt{2\pi\rho}\frac{\sigma_v}{\sqrt{\sigma_\phi}}
\end{equation}
where the $\rho$ is the density of ISM, the $\sigma_v$ as velocity dispersion relates to turbulence, $\sigma_\phi$ is the magnetic field orientation angle dispersion which is affected by small-scale turbulence.
The assumption of ST method \citep{2021A&A...647A.186S,2021AA...647A..78A} is similar to the DCF method \citep{1951PhRv...81..890D,1953ApJ...118..113C,2004ApJ...600..279C}.
Different occur in the handle of angle dispersion $\sigma_\phi$, where that in ST method is $\sigma_\phi^{-0.5}$ and in DCF method is $\sigma_\phi^{-1}$ (DCF method : $B_{\rm pos} = Q\sqrt{4\pi\rho}\frac{\sigma_v}{\sigma_\phi}$).
Regarding error propagation, the ST method could reduce the error from angle dispersion to magnetic field strength compared to the DCF method.
Thus, we use the ST to estimate the plane distribution of magnetic field strength in OMC-1.

As Fig.\,\ref{figD1} shows, we calculate the basic parameters, density $\rho$, non-thermal velocity dispersion $\sigma_{\rm v, NT}$, and magnetic field orientation angle dispersion in the same scale ($\sim$ 40$''$, 0.07 pc).
The volume density comes from the column density observed by Herschel \citep{2010A&A...518L...2P,2010A&A...518L...3G,2013ApJ...763...55R,2013ApJ...777L..33P} (see Ap.\,\ref{adata}), using the Constrained Diffusion Method \citep{2022ApJS..259...59L} and its derivative method \citep{2025arXiv250801130Z}
The angle dispersion is calculated by magnetic field orientation derived by 850$\mu$m dust polarization \citep{2017ApJ...846..122P} and mapped by the sub-block of 3$\times$3 beams \citep{2025arXiv250801130Z}.

The non-thermal velocity dispersion comes from the velocity dispersion and kinetic temperature of NH$_3$ observed by GBT \citep{2017ApJ...843...63F}.

The distribution of total magnetic field strength in OMC-1 is estimated by POS component of magnetic field strength \citep{2004ApJ...600..279C}:
\begin{equation}
    B_{\rm tot} = \frac{4}{\pi}B_{\rm pos}
\end{equation}
The distribution is shown in Fig.\,\ref{figD1}

\end{document}